%% file: sigir26-sr4all-frame.tex
\documentclass[sigconf,natbib=true]{acmart}
\usepackage{sigir26-sr4all-frame}

\begin{document}
\input{sigir26-sr4all-pre}

\input{sigir26-sr4all-part1}
\input{sigir26-sr4all-part2}
\input{sigir26-sr4all-part3}

\input{sigir26-sr4all-part4}
\input{sigir26-sr4all-part5}
\input{sigir26-sr4all-sum}

\bibliographystyle{ACM-Reference-Format}
\bibliography{sigir-26-sr4all-lit} 

\end{document}

%% file: sigir26-sr4all-pre.tex
\title{Webis-SR4ALL-26: A Large-Scale Multi-Domain Corpus for Systematic Review Retrieval and Screening}
\title{A Large-Scale Corpus of Systematic Reviews in All Sciences}
\title{A Large-Scale Multi-Domain Corpus of Systematic Reviews}
\title{A Large-Scale, Cross-Disciplinary Corpus of Systematic Reviews}

\settopmatter{authorsperrow=5}

\author{Pierre Achkar}
\affiliation{%
  \institution{Leipzig University; Fraunhofer ISI}
  \country{}
}

\author{Tim Gollub}
\affiliation{%
  \institution{Bauhaus-Universität Weimar}
  \country{}
}

\author{Arno Simons}
\affiliation{%
  \institution{\mbox{TU Berlin}}
  \country{}
}

\author{Harrisen Scells}
\affiliation{%
  \institution{University of Tübingen}
  \country{}
}

\author{Martin Potthast}
\affiliation{%
  \institution{Kassel University; hessian.AI; ScaDS.AI}
  \country{}
}

\begin{abstract}
Existing benchmarks for systematic reviewing remain limited either in scale or in disciplinary coverage, with some collections comprising only a modest number of topics and others focusing primarily on biomedical research. We present Webis-SR4ALL-26, a large-scale, cross-disciplinary corpus of 301,871~systematic reviews spanning all scientific fields as covered by OpenAlex.
Using a multi-stage pre-processing pipeline, we link reviews to resolved OpenAlex metadata and reference lists and extract, when explicitly reported, structured method artifacts relevant to retrieval and screening. 
These artifacts include reported search strategies (Boolean queries or keyword lists) that we normalize into executable approximations, as well as reported inclusion and exclusion criteria. Together, these layers support cross-domain benchmarking of retrieval and screening components against review reference lists, training and evaluation of extraction methods for review artifacts, and comparative meta-science analyses of systematic review practices across disciplines and time. To demonstrate one concrete use case, we report large-scale baseline retrieval signals by executing normalized search strategies in OpenAlex and comparing retrieved sets to resolved reference lists. We release the corpus and the pre-processing pipeline, along with code used for extraction validation and the retrieval demonstration.%
\footnote{%
\url{https://github.com/webis-de/sigir26-sr4all}, \url{https://doi.org/10.5281/zenodo.18431942}}
\end{abstract}

\ccsdesc[500]{Information systems~Information retrieval}

\keywords{Systematic reviews, Cross-Disciplinary, Retrieval, Screening}

\maketitle

%% file: sigir26-sr4all-part1.tex
\section{Introduction}

Systematic reviews have become a dominant method and, in many areas, the standard for synthesizing evidence to answer focused research questions, especially in medicine, education, and the social sciences. Systematic reviews follow a protocolled and transparent approach to searching for, selecting, and documenting relevant literature, with explicit inclusion and exclusion criteria. While they share this systematic approach to evidence identification and screening, they can differ substantially with respect to synthesis methods, ranging from qualitative evidence syntheses and mixed-methods reviews to quantitative meta-analyses that aggregate effect sizes~\cite{gough:2012}. Systematic reviews are valued for their completeness in capturing the available evidence, transparency in documenting procedures and data sources, reproducibility through well-defined methodologies, rigorous appraisal of study quality (including risk-of-bias assessment and related threats to validity), and their ability to support more robust and informative conclusions via structured synthesis across studies~\cite{liberati:2009, mulrow:1994}. Their results guide subsequent research, policy, and professional practice.

Historically, systematic review methodology was developed and institutionalized most prominently in biomedicine to extract and communicate ``the evidence'' from an expanding primary literature for decision-making~\cite{chalmers:1993, guyatt:1992}. Related evidence-based policy and practice agendas then helped diffuse evidence-synthesis toolkits beyond health, supported by new institutions and infrastructures for producing and disseminating reviews~\cite{simons:2021}. Over time, these principles spread into many other disciplines, often with field-specific adaptations in information sources, search conventions, eligibility criteria, and synthesis approaches. Regardless of the domain and application, a systematic review is typically conducted in a multi-step workflow, from scoping the objective and inclusion criteria, through searching for and selecting studies (retrieval and screening), collecting and appraising study data, and finally synthesizing, presenting, and interpreting findings. Methodological guides in biomedicine~\cite{chandler_cochrane_2019}, nursing~\cite{bettany_saltikov_how_2012}, education~\cite{newman_systematic_2020}, social science~\cite{petticrew_systematic_2008}, design~\cite{lame:2019}, or computer science~\cite{kitchenham:2004} cover these core steps, while differing in granularity and how they group them into phases.

Conducting a systematic review is time-consuming and expensive. Consequently, substantial effort has been devoted to automating one of the most expensive parts of the process, namely retrieval and screening. However, existing evaluation resources for retrieval and screening remain largely concentrated in the biomedical domain (see Table~\ref{table-sr-datasets}). This focus reflects the empirical dominance of systematic reviews in biomedicine, but constrains evaluation in other scientific fields. Consequently, domain-specific query formulations, search strategies, and relevance criteria outside biomedicine remain sparsely represented in current collections, limiting the assessment of retrieval and screening methods beyond the medical domain.

\input{table-sr-datasets}

To address this gap, we introduce Webis-SR4ALL-26, a large-scale, cross-disciplinary corpus of 301,871~systematic reviews spanning 27~scientific domains, derived from reviews indexed in OpenAlex~\cite{Priem2022OpenAlexAF}. A large subset of reviews is linked to its resolved set of cited studies, providing a stable bibliographic context. In addition, a broad range of methodological information is extracted from the available review full texts when explicitly reported. For systematic reviews that report a search strategy, whether as explicit Boolean queries or keyword lists, we construct a canonical Boolean representation as a minimal approximation of the reported strategy for execution in OpenAlex. Together, these components make Webis-SR4ALL-26 a rich multi-layer resource for retrieval and screening in systematic reviews across domains and review types, and for multiple research communities. Its layers (links, method fields, normalized queries) can each serve as supervision, evaluation targets, or observational signals for descriptive analyses. The corpus supports benchmarking and development of retrieval, screening, and extraction approaches, and enables comparative meta-research, including science-of-science and sociological studies of how query formulation and reporting practices evolve across domains and review styles over time.

%% file: table-sr-datasets.tex
\newcommand{\bsq}{$\blacksquare$}
\newcommand{\wsq}{$\square$}

\begin{table*}[t]
\small
\sffamily
\renewcommand{\arraystretch}{1.05}
\renewcommand{\tabcolsep}{3.6pt}
\centering
\caption{Overview of systematic review datasets.}
\label{table-sr-datasets}
\vskip-2ex
\begin{tabular}{@{}l@{}rcc@{\hspace{2em}}lcccc@{\hspace{2em}}rcccccc@{\hspace{2em}}ccccc@{}}
\toprule
  \multicolumn{4}{@{}c@{\hspace{2em}}}{\bfseries Dataset}                                                                                                   & \multicolumn{5}{@{}c@{\hspace{2em}}}{\bfseries Publication Data} & \multicolumn{12}{@{}c@{}}{\bfseries Structured Data Fields}    \\
\cmidrule(r@{2em}){1-4}\cmidrule(r@{2em}){5-9}\cmidrule{10-21}
  Name             & \multicolumn{1}{@{}c@{}}{Rf.} & Year               & Link                                         & Domain & \multicolumn{3}{@{}c@{}}{\kern-0.5em SysRevs} & Pool & Topics & \multicolumn{6}{@{}c@{\hspace{2em}}}{Topic Description} & \multicolumn{5}{@{}c@{}}{Screening} \\
\cmidrule(r@{\tabcolsep}){1-1}\cmidrule(r@{\tabcolsep}){2-2}\cmidrule(l@{\tabcolsep}r@{\tabcolsep}){3-3}\cmidrule(l@{\tabcolsep}r@{2em}){4-4}\cmidrule(r@{\tabcolsep}){5-5}\cmidrule(l@{\tabcolsep}r@{\tabcolsep}){6-8}\cmidrule(l@{\tabcolsep}r@{2em}){9-9}\cmidrule(r@{\tabcolsep}){10-10}\cmidrule(l@{\tabcolsep}r@{2em}){11-16}\cmidrule{17-21}
                   &                        &                           &                                                                                       &         & Rfs. &  TA  &  FT  &          &        & Obj. & RQs  & KWs  & BQs  & Crt. & Rng. & DBs  &  CR  &  CI  &  PR  &  PI  \\
\midrule
  Cohen et al.     & \cite{cohen:2006}      & \citeyear{cohen:2006}     & \href{https://dmice.ohsu.edu/cohenaa/systematic-drug-class-review-data.html}{\faLink} & Biomed  &      &      &      &          &     15 &      &      &      &      &      &      &      & \bsq & \bsq & \bsq & \bsq \\
  Wallace et al.   & \cite{wallace:2010}    & \citeyear{wallace:2010}   & \href{https://github.com/bwallace/citation-screening}{\faLink}                        & Biomed  &   \bsq   &      &      &          &      3 &      &      &      &      &      &      &      & \bsq & \bsq &      &      \\
  SWIFT-Review     & \cite{howard:2016}     & \citeyear{howard:2016}    & \href{https://link.springer.com/article/10.1186/s13643-016-0263-z#Sec30}{\faLink}     & Biomed  &      &      &      &          &      5 &      &      &      &      &      &      &      & \bsq & \bsq & \bsq & \bsq \\
  SysRev Query C.  & \ \cite{scells:2017}   & \citeyear{scells:2017}    & \href{https://github.com/ielab/SIGIR2017-SysRev-Collection}{\faLink}                  & Biomed  &   \bsq   &      &      &     PubMed     &     94 &      &      &      & \bsq &      & \bsq &      & \bsq & \bsq & \bsq & \bsq \\
  CLEF TAR 2017    & \cite{kanoulas:2017}   & \citeyear{kanoulas:2017}  & \href{https://github.com/CLEF-TAR/tar}{\faLink}                                       & Biomed  &   \bsq    &  \wsq    &      &      PubMed    &     50 &      &      &      & \bsq &      & \bsq &      & \bsq & \bsq & \bsq & \bsq \\
  CLEF TAR 2018    & \cite{kanoulas:2018}   & \citeyear{kanoulas:2018}  & \href{https://github.com/CLEF-TAR/tar}{\faLink}                                       & Biomed  &  \bsq     &   \wsq   &      &      PubMed    &     30 &      &      &      & \bsq &      & \bsq &      & \bsq & \bsq & \bsq & \bsq \\
  CLEF TAR 2019    & \cite{kanoulas:2019}   & \citeyear{kanoulas:2019}  & \href{https://github.com/CLEF-TAR/tar}{\faLink}                                       & Biomed  &   \bsq    &   \wsq   &      &     PubMed     &     49 &      &      &      & \bsq &      & \bsq &      & \bsq & \bsq & \bsq & \bsq \\
  Alharbi et al.   & \cite{alharbi:2019}    & \citeyear{alharbi:2019}   & \href{https://github.com/Amal-Alharbi/Systematic_Reviews_Update}{\faLink}             & Biomed  &  \bsq   &  \wsq    &      &      PubMed   &     25 &      &      &   \bsq   &   \bsq   &      &      &      &      &   \bsq   &      &   \bsq   \\
  Hannousse et al. & \cite{hannousse:2021}  & \citeyear{hannousse:2021} & \href{https://github.com/hannousse/Semantic-Scholar-Evaluation}{\faLink}              & CompSci &  \bsq    &      &      &     S2     &      7 &      &      &      &      &      &      &      &      &  \bsq    &      &  \bsq    \\
  SysRev Seed C.   & \cite{wang:2022}       & \citeyear{wang:2022}      & \href{https://github.com/ielab/sysrev-seed-collection}{\faLink}                       & Biomed  &   \bsq   &  \wsq    &      &     PubMed     &     40 & \bsq &      &      & \bsq &      & \bsq &      & \bsq & \bsq & \bsq & \bsq \\
  CSMeD            & \cite{kusa:2023}       & \citeyear{kusa:2023}      &  \href{https://github.com/WojciechKusa/systematic-review-datasets}{\faLink}            & Both    &   \bsq   &  \wsq    &      &   Mixed  &    325 & \bsq &      &      & \bsq & \bsq &      & \bsq & \bsq & \bsq & \bsq & \bsq \\
  AutoBool         & \cite{wang:2025b}      & \citeyear{wang:2025b}     & \href{https://github.com/ielab/AutoBool}{\faLink}                                     & Biomed  &   \bsq   &  \wsq    &      &     PubMed     & 65,588 &      &      &      &      &      & \bsq &      &      & \bsq &      & \bsq \\
\midrule
  \multicolumn{2}{@{}l@{}}{Webis-SR4ALL-26} & 2026                      & \href{https://doi.org/10.5281/zenodo.18431943}{\faLink}                               & Open    & \bsq & \bsq & \wsq & OpenAlex & 301,871 & \wsq & \wsq & \wsq & \wsq & \wsq & \wsq & \wsq & \wsq & \wsq &      & \bsq \\
\bottomrule
\addlinespace
\multicolumn{21}{@{}l@{}}{
\parbox{\textwidth}{
\footnotesize Abbreviations: For systematic reviews (SysRevs), we distinguish whether they are available as references (Rfs.), title and abstract (TA), and full texts (FT). For the structured data fields, we distinguish objective~(Obj.), research questions~(RQs), keywords~(KWs), Boolean queries~(BQs), inclusion and exclusion criteria~(Crt.), publication date ranges~(Rng.), publication databases originally queried~(DBs), counts of retrieved~(CR) and included~(CI) publications, and references to publications retrieved~(PR) and included~(PI) in the systematic review. Filled boxes indicate full availability; empty boxes indicate partial availability.}}
\end{tabular}
\end{table*}

%% file: sigir26-sr4all-part2.tex
\section{Related Work}

Existing information retrieval test collections for evaluating retrieval and screening in systematic reviewing have predominantly focused on the biomedical domain. While these datasets differ in scale, annotation depth, and supported tasks, they share a common goal: providing reusable test collections for studying search (i.e., retrieving literature), screening (i.e., determining if literature should be included in the review or not), and study prioritization (i.e., ranking the literature to assist with the screening phase) in systematic reviews. Table~\ref{table-sr-datasets} summarizes the most commonly used resources and contrasts them with Webis-SR4ALL-26.

We first discuss typical test collections that are commonly used in systematic review automation tasks. The SIGIR~2017 SysRev Query Collection~\cite{scells:2017} comprises Cochrane reviews published between~2014 and~2016. The original Boolean search strategies were translated into executable Elasticsearch queries. Included and excluded studies were mapped to PubMed identifiers. The Seed Studies Collection~\cite{wang:2022} targets seed-driven search strategies in systematic reviewing; i.e., the creation of Boolean queries driven by knowledge about a priori studies that are likely to be included in the review. It provides rich review-process artifacts, including original PubMed Boolean queries, seed studies used to assist in the formulation of the Boolean queries, and lists of included studies. In addition, the collection includes references obtained through automated citation chasing tools such as CitationChaser and SpiderCite. The CLEF TAR collections~\cite{kanoulas:2017,kanoulas:2018,kanoulas:2019} were released as part of the CLEF lab on technology-assisted reviews in empirical medicine. Built on Cochrane reviews, it contains expert-formulated queries, retrieved records, and relevance assessments spanning intervention, prognosis, qualitative studies, and diagnostic test accuracy review types. The corpus primarily supports querying and screening prioritization tasks. 

Apart from the typical retrieval and screening tasks, \citet{alharbi:2019} proposed a dataset for performing these tasks on systematic review updates, where there are two versions of the review and the task is to assist in helping identify relevant studies for the updated version. The dataset from~\citet{hannousse:2021} is used exclusively for screening and is the only other non-biomedical systematic review collection we identified in our review of the literature. The CSMeD dataset~\cite{kusa:2023} consolidated prior biomedical and computer science screening datasets~\cite{cohen:2006,wallace:2010,howard:2016,scells:2017,kanoulas:2017,kanoulas:2018,kanoulas:2019,alharbi:2019,hannousse:2021} into a unified meta-collection of 325~systematic reviews. By harmonizing nine existing resources and providing standardized access via the BigBio framework, CSMeD enables more comparable evaluations of citation screening methods. The dataset further enriches a subset of Cochrane reviews by including eligibility criteria and search strategies extracted from review protocols, as well as CSMeD-FT, a full-text screening collection with time-based splits to prevent data leakage. Finally, the most recent dataset for systematic review automation we identified was from \citet{wang:2025b}. This dataset was used primarily as training data for Boolean query formulation through reinforcement learning. While it is by far the largest dataset compared to all others discussed above, it provides fewer review-process artifacts and is not as useful for many other tasks for automating systematic reviews.

While these resources have been instrumental in advancing retrieval and screening methods for biomedical systematic reviews, their scope usually remains limited to a single scientific domain and a small number of curated review topics. In contrast, Webis-SR4ALL-26 is desigend as a large-scale, cross-disciplinary corpus covering hundreds of thousands of systematic reviews from a variety of scientific fields. It enables research into systematic reviews at a scale previously not possible, and across tasks like retrieval and screening. With further extraction and annotation effort, it may also support future tasks like syntehsizing systematic reviews.

%% file: sigir26-sr4all-part3.tex
\section{The Webis-SR4ALL-26 Corpus}

The corpus is designed as a large-scale, cross-disciplinary collection of systematic reviews to support research on evidence synthesis workflows, with a particular focus on the retrieval and screening stages. The construction follows a three-stage enrichment pipeline: (i) large-scale data collection and metadata filtering; (ii) full-text acquisition and document parsing; and (iii) structured information extraction. Inclusion is determined at the metadata level, with subsequent stages augmenting eligible records with full-text representations and structured information. 

\subsection{Data Collection and Filtering}
Candidate systematic reviews were retrieved from the OpenAlex database on September~30,~2025. We queried the \texttt{title} field for the phrases ``systematic review'' and ``systematic literature review'', restricting results to works of type \texttt{review}. This retrieval yielded 485,446 candidate records.

To remove duplicates arising from multiple identifiers or minor title variations, records were deduplicated using a combination of DOI, OpenAlex ID, and normalized title strings. This process identified and removed 20,343 duplicate entries.

We then applied metadata-based inclusion filters to ensure bibliographic consistency and cross-source integration. Records were excluded if their OpenAlex reference list was empty, if they were not written in English, or if they lacked a valid DOI. In addition, we applied title-based inclusion heuristics to isolate explicitly declared systematic reviews and exclude loosely defined review types. Titles were required to contain an explicit formulation of ``systematic review'' (optionally ``systematic literature review''), including common syntactic variants (e.g., of, in, on, or a colon). To focus on original review studies, we excluded works whose titles indicated updates or revisions, such as explicit mentions of ``update'' or ``updated systematic review''. After applying these filters, 280,886 unique systematic reviews remained in the OpenAlex-derived metadata layer.

To extend coverage beyond OpenAlex title-based retrieval, we incorporated systematic reviews from established benchmark datasets, including CLEF TAR 2017--2019~\cite{kanoulas:2017,kanoulas:2018,kanoulas:2019}, SysRev-Query~\cite{scells:2017}, SysRev-Seed~\cite{wang:2022}, CSMed~\cite{kusa:2023}, and AutoBool~\cite{wang:2025b}. After internal deduplication across these resources, 65,906 unique reviews were obtained. DOI-based matching against the OpenAlex-derived metadata layer showed that 44,921 were already covered. For the remaining 20,985 unmatched reviews, we resolved their OpenAlex identifiers and incorporated the corresponding records into the metadata layer. 

After incorporating the unmatched benchmark reviews, the Webis-SR4ALL-26 metadata layer comprises 301,871 systematic reviews in total. All records are linked to OpenAlex identifiers, providing a unified metadata and citation backbone. Table~\ref{tab:sr4all-metadata-filtering} summarizes the sequential construction of the metadata layer.

To characterize the cross-disciplinary composition of the corpus, we rely on the OpenAlex primary field assignment for each work. The corpus spans 27 primary fields in total, covering a broad range of scientific domains. Medicine accounts for approximately 64\% of reviews. The largest non-medical fields are Psychology (6\%), Health Professions (5\%), Biochemistry, Genetics and Molecular Biology (4\%), Social Sciences (3\%), and Neuroscience (3\%). Dentistry, Environmental Science, Engineering, Computer Science, Immunology and Microbiology, Economics, and related disciplines, which contribute smaller shares. This distribution indicates broad cross-domain coverage, while also highlighting differences in how frequently systematic review methodology is used across disciplines.

In addition to standardized metadata, OpenAlex provides indexed reference lists that form the basis for reference resolution in our setup. All citation links are derived exclusively from OpenAlex, with no external reference parsing, matching or enrichment. As an open, cross-domain infrastructure with broad coverage of scholarly works and citation relations, OpenAlex thus provides a consistent and reproducible foundation for large-scale corpus construction and citation graph derivation~\cite{culbert2025openalexrefcoverage}.

\input{table-filtering-stats}

\subsection{Full-Text Acquisition and Document Parsing}
We assessed full-text availability for all records in the metadata layer using the PDF links provided in OpenAlex. Where a PDF link was available, we attempted automated download of the full text. This process yielded 72,678 successfully retrieved PDFs. Records without a PDF link, or with non-functional links, remain in the corpus without the full text being available.

All valid PDF documents were parsed into markdown using \textit{PaddleOCR-VL},%
\footnote{\url{https://github.com/PaddlePaddle/PaddleOCR}}
a 0.9B-parameter vision--language model for multilingual document parsing. The model is designed to process complex scientific layouts, including multi-column text, tables, and formulas, while maintaining relatively low computational requirements. On the olmOCR-Bench benchmark,%
\footnote{\url{https://github.com/allenai/olmocr/tree/main/olmocr/bench}}
PaddleOCR-VL achieves competitive overall effectiveness compared to substantially larger OCR systems, motivating its selection as a balanced trade-off between accuracy and efficiency.

\subsection{Structured Information Extraction}
\begin{figure*}[t]
    \centering
    \includegraphics[width=\linewidth]{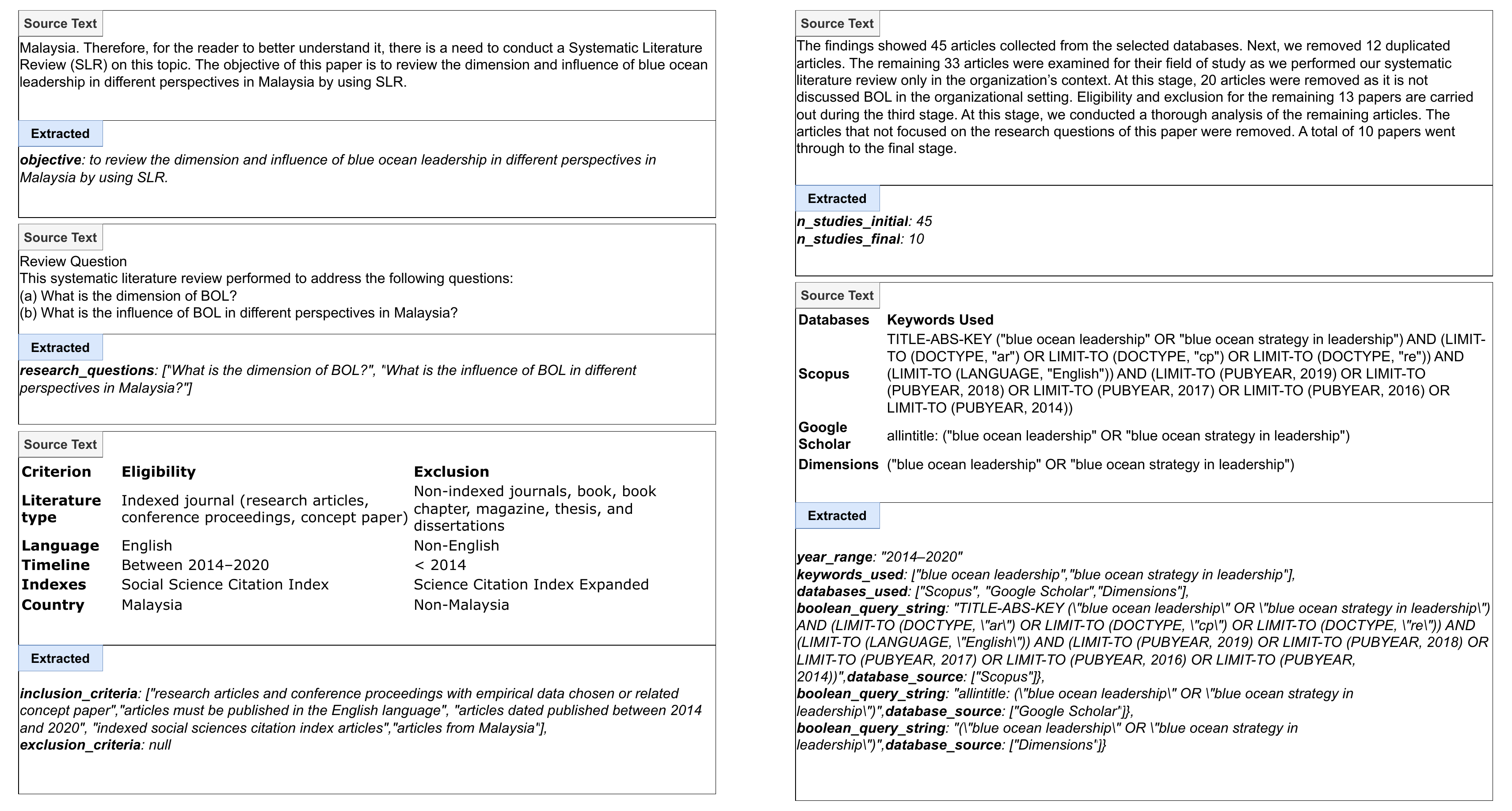}
    \caption{Grounded extraction of review key information. Verbatim source spans from the PDF are transformed into structured schema fields via the verify-then-repair pipeline.}
    \label{fig:extraction_example}
\end{figure*}

To enrich the corpus with structured metadata, we apply an automated LLM-based extraction pipeline to the OCR-parsed full texts. As illustrated in Figure~\ref{fig:extraction_example}, the pipeline maps raw source content to our structured schema. Given the scale and heterogeneity of the corpus, the pipeline prioritizes precision, verifiability, and transparency over completeness, as manual curation does not scale to tens of thousands of reviews.

The pipeline targets methodological fields commonly reported in systematic reviews: study objective, research questions, search strategies (Boolean queries or keywords), eligibility criteria, numbers of studies identified and included, temporal search restrictions, databases queried, and citation chasing strategies. 

Search strategies are of particular importance, as they provide a basis for approximating expert-written queries. Extracted Boolean expressions or keyword formulations can be normalized into executable queries for retrieval experiments, enabling controlled comparisons between author-designed strategies and alternative query generation methods. Study objectives and eligibility criteria further contextualize these strategies and support structured evaluation settings. Reviews with search strategies can therefore be used for direct query-based retrieval experiments. Reviews lacking explicit queries remain valuable for evaluation through their curated reference sets, which serve as relevance judgments independent of the reported search formulation.

\paragraph{Verify-Then-Repair Pipeline}
To prevent hallucinations and ensure extraction quality, we employ a grounded extraction strategy with iterative repair, based on prior work showing improved extraction accuracy through evidence anchoring~\cite{li:2025}. In the \textbf{Primary Pass}, the Qwen3-32B model extracts all schema-defined fields from the full text, generating candidate values paired with verbatim evidence spans copied directly from the source document. Each extraction then undergoes a two-stage verification process. First, candidates are validated through approximate string matching against the OCR text to ensure the evidence span exists in the document, tolerating minor parsing noise~\cite{barrow:2025}. Second, MiniCheck, an LLM-based verifier, assesses whether the extracted value is factually supported by the provided evidence span~\cite{tang:2024}. Any extraction failing either alignment or semantic verification is strictly discarded and reset to \texttt{null}. To maximize recall without compromising precision, a \textbf{Repair Pass} is triggered for fields that remain \texttt{null} after the primary pass. The model receives a new, targeted prompt focusing only on the missing fields while leaving the successfully extracted values unchanged. The repair outputs undergo the same two-stage verification protocol (\textbf{Final Validation}), and only extractions that survive all verification checks are included in the final corpus.

The final output is an evidence-centric structured representation integrating only verified extractions with OpenAlex metadata. For comparability, temporal search constraints are normalized to a canonical year-range format anchored to the publication year, while preserving original values. Table~\ref{tab:sysrev_field_coverage} summarizes metadata and methodological field coverage across the corpus after enrichment.

\input{table_field_cov}

\subsection{Qualitative Evaluation of Extraction}
To assess extraction quality, we conducted a qualitative analysis of 60 randomly sampled reviews across three outcome categories: mostly filled (n=20), partially filled (n=20), and mostly null (n=20). For each review, we manually compared the original PDF with the extracted JSON to determine whether target fields were present in the source, correctly extracted, incorrectly extracted, or filtered during verification. This allowed us to characterize field-level performance and identify systematic failure modes.

Extraction performance was strongest in the mostly filled subset. In these cases, the pipeline reliably captured Boolean queries, study counts, eligibility criteria, temporal search ranges, and study objectives. Successful extractions were typically associated with clear document structure, explicit section headings (e.g., Search Strategy, Eligibility Criteria), and standardized reporting. This suggests that the pipeline performs robustly when methodological information is clearly outlined and positioned within the document.

A different pattern emerged in the partially filled subset. In many cases, the model correctly identified the relevant information during the initial extraction stage, but subsequent verification stages nullified these fields. Information distributed across multiple sentences or requiring light abstraction was more frequently rejected under strict evidence-matching constraints. For instance, eligibility criteria, snowballing indicators, and initial study counts appeared in the raw extraction trace yet were absent from the final structured output. Overall, verification filtering primarily affected fields not explicitly stated in a single contiguous span, reflecting a focus on high precision for clearly stated fields with lower recall for information spread across the document.

Most failures in the null subset were attributable to document type misalignment rather than extraction errors. The majority of the sampled records corresponded to errata, supplementary materials, or registration documents that did not contain systematic review methodology. In these cases, the absence of extracted fields reflected the absence of target information in the source text.

%% file: table-filtering-stats.tex
\begin{table}[t]
\centering
\sffamily
\renewcommand{\arraystretch}{1.2}
\caption{Statistics of the metadata-layer construction process.}
\begin{tabular*}{\columnwidth}{@{\extracolsep{\fill}}lrr}
\toprule
\textbf{Construction Pipeline} & \textbf{Remaining} & $\Delta$ \\
\midrule
OpenAlex title search & 485{,}446 & -- \\
1. Deduplication & 465{,}103 & -4.2\% \\
2. Title heuristic & 293{,}523 & -36.9\% \\
3. Not English & 287{,}061 & -2.2\% \\
4. No DOI & 283{,}560 & -1.2\% \\
5. Is Update & 280{,}886 & -0.9\% \\
\midrule
+ Benchmark integration & 301{,}871 & +7.5\% \\
\bottomrule
\end{tabular*}
\label{tab:sr4all-metadata-filtering}
\end{table}

%% file: table_field_cov.tex
\begin{table}[t]
\centering
\sffamily
\renewcommand{\arraystretch}{1.2}
\caption{Availability of fields across Webis-SR4ALL-26. 
The effectively usable subset comprises reviews containing an objective, at least one search strategy representation (Boolean query or keyword list), and eligibility criteria.}

\begin{tabular*}{\columnwidth}{@{\extracolsep{\fill}}lrr}
\toprule
\textbf{Subset} & \textbf{Count} & \textbf{Percentage} \\
\midrule
Total available sysrevs & 301,871 & 100\% \\
\midrule
\multicolumn{3}{l}{\textit{Availability of core materials}} \\
w/ title & 301,871 & 100.0\% \\
w/ abstract & 227,208 & 75.3\% \\
w/ full texts & 72,678 & 24.1\% \\
w/ all in this group & 54,812 & 18.2\% \\
\midrule
\multicolumn{3}{l}{\textit{Review methodology fields}} \\
w/ objective & 59,108 & 19.6\% \\
w/ research questions & 14,934 & 4.9\% \\
w/ keywords used & 36,028 & 11.9\% \\
w/ boolean queries & 28,251 & 9.4\% \\
w/ inclusion criteria & 48,010 & 15.9\% \\
w/ exclusion criteria & 24,914 & 8.3\% \\
w/ date range & 61,951 & 20.5\% \\
w/ all in this group & 3,674 & 1.2\% \\
\midrule
\multicolumn{3}{l}{\textit{Search and study flow data}} \\
w/ databases & 9,283 & 3.1\% \\
w/ paper counts retrieved & 46,590 & 15.4\% \\
w/ paper counts included & 51,595 & 17.1\% \\
w/ paper references included & 301,871 & 100.0\% \\
w/ all in this group & 6,379 & 2.1\% \\
\midrule
w/ all of the above & 733 & 0.2\% \\
\midrule
\textbf{Effectively usable subset} & \textbf{38,292} & \textbf{12.7\%} \\
\bottomrule
\end{tabular*}
\label{tab:sysrev_field_coverage}
\end{table}

%% file: sigir26-sr4all-part4.tex

\section{Intended Use and Retrieval Experiments}
We first summarize intended uses of Webis-SR4ALL-26. We then present a retrieval demonstration by executing a subset of normalized search strategies in OpenAlex and analyzing the resulting retrieval signals.

\subsection{Intended Use}
Webis-SR4ALL-26 enables large-scale evaluation and analysis of systematic review retrieval and screening across scientific domains. Key applications include: 
(i) benchmarking retrieval and screening methods against the review’s reference list; 
(ii) analyzing cross-domain differences in retrieval and screening performance; 
(iii) training and evaluating extraction methods for systematic review artifacts;
and (iv) studying systematic review practices across fields and over time.

\subsection{Query Normalization and Demonstration Execution}
\label{sec:query-norm-demo}
To enable controlled experimentation, reported search strategies are transformed into a uniform representation that can be executed within a single retrieval environment. As these strategies are expressed in heterogeneous, database-specific query languages, direct large-scale execution is not feasible. We therefore convert them into simplified Boolean expressions that can be issued to OpenAlex via its API. The resulting normalized queries are included in Webis-SR4ALL-26 and released as part of the dataset to facilitate reproducible retrieval experiments.

The normalization process retains only topical terms and the fundamental logical operators \texttt{AND}, \texttt{OR}, and \texttt{NOT}. Database-specific syntax has been removed, including explicit field tags (e.g. \texttt{TITLE-ABS-KEY} in Scopus or \texttt{[mesh]} in Pubmed) and metadata constraints, such as language or publication type filters. Temporal restrictions are not embedded in the Boolean string; rather, extracted year-range metadata is applied externally at query time

Normalization is performed using an instruction-tuned large language model (Qwen3-32B) applied to reported Boolean queries or keyword lists. The model is prompted using a few-shot approach to preserve the topical vocabulary and high-level logical structure of each strategy without introducing novel concepts or search terms not present in the original formulation. Lightweight rule-based post-processing then ensures syntactic validity of the resulting Boolean expressions.

Some strategies cannot be normalized without additional assumptions. For instance, when numeric placeholders reference external term lists not provided in the review, or when the reported strategy contains insufficient information to construct a meaningful query, the normalized query is set to \texttt{null}. Similarly, when post-processing cannot produce a valid, non-empty Boolean expression, the query is also set to \texttt{null}. 

All queries that could be executed against the OpenAlex \texttt{/works} endpoint were first partitioned into three buckets based on their initial result counts returned via the \texttt{search=}. The buckets were defined using fixed hit-count thresholds: (i) 1,000–250,000 results, (ii) 250,000–1,000,000 results, and (iii) more than 1,000,000 results. The analysis was restricted to the smallest bucket.

The executed subset comprises two query variants derived from the review data: (i) normalized Boolean strategies reconstructed from the reported search strings, and (ii) keyword-based queries constructed from reported keyword lists in cases where full Boolean queries were unavailable.

For each executed query, retrieved OpenAlex work identifiers were aligned with the corresponding review’s reference set. This alignment provides per-review retrieval signals, including precision-, recall-, and F-score–based metrics quantifying the extent to which the strategy recovers its cited studies. Aggregate results over this subset are reported in Table~\ref{tab:retrieval-aggregate}.

\input{table-agg-res}

%% file: table-agg-res.tex
\begin{table}[t]
\centering
\sffamily
\renewcommand{\arraystretch}{1.1}
\caption{Summary retrieval metrics for proof-of-concept query execution.}
\label{tab:retrieval-aggregate}
\begin{tabular*}{\columnwidth}{@{\extracolsep{\fill}}llllll}
\toprule
\textbf{Query origin} & \textbf{\# Reviews} & \textbf{Prec.} & \textbf{Recall} & \textbf{F$_1$} & \textbf{F$_3$} \\
\midrule
Boolean queries & 12,005 & 0.014 & 0.245 & 0.019 & 0.050 \\
Keyword lists   & 9,043 & 0.016 & 0.180 & 0.018 & 0.043 \\
\bottomrule
\end{tabular*}
\end{table}

%% file: sigir26-sr4all-part5.tex
\section{Discussion}

Webis-SR4ALL-26 provides a large-scale, cross-disciplinary corpus of systematic reviews with a uniform, fully open metadata backbone. By anchoring reviews and cited works to OpenAlex and by extracting method-related fields from available full texts, the corpus combines bibliographic context with structured descriptions of review objectives, eligibility criteria, and reported search strategies. This design prioritizes verifiability and cross-domain comparability, making it possible to study systematic review workflows and practices under consistent conditions across disciplines, while accepting uneven coverage due to incomplete full-text availability and heterogeneous reporting.

To illustrate one concrete use case, Section~\ref{sec:query-norm-demo} reported retrieval signals in a controlled setting using the executable query layer. Extracted strategies were normalized into a canonical Boolean form and executed against OpenAlex where feasible, with retrieved results compared to resolved reference lists. In this demonstration, normalized Boolean strategies achieved higher recall than keyword-based queries (0.245 vs.\ 0.180) at similarly low precision (0.015). These signals do not aim to reproduce database-specific search performance. They should be interpreted as retrieval behaviour within a simplified, uniform index that is suitable for cross-domain comparison and method development.

Beyond this retrieval demonstration, the corpus supports use cases that do not depend on executable queries. For screening research, the reference backbone and method fields support the study and benchmarking of prioritization and selection components in systematic review pipelines, while making explicit that reference lists are not identical to included-study sets. For extraction research, the structured fields derived from full text can serve as targets for models that identify and recover review artifacts such as objectives, eligibility criteria, and search strategies. For meta-research and science studies, Webis-SR4ALL-26 enables comparative analyses of systematic review practices across disciplines and time, and supports investigations into the often neglected role of systematic reviews in science communication and public policy~\cite{blumel:2020, mcmahan:2021, schniedermann:2022, simons:2021}.

Several limitations follow from the design choices underlying the corpus construction. The identification of systematic reviews relies on explicit title-based declarations, which favors precision but excludes reviews that do not self-identify using this terminology, which is a known challenge~\cite{moher:2007}. Full-text availability is incomplete, which constrains methodological extraction to a subset of records and contributes to uneven field coverage. In addition, the strict verification protocol applied during extraction prioritizes precision over recall, so methodological elements that are implicitly stated or distributed across multiple passages may remain unstructured. The resolved reference sets used for evaluation do not correspond directly to included-study gold standards, as reviews cite background and contextual literature in addition to eligible studies. Finally, the normalization of reported search strategies into simplified executable representations introduces abstractions that may not fully capture database-specific syntax, field restrictions, or manual refinements present in the original formulations.

Future work can address several of these constraints. Systematic review identification could be extended beyond title-based heuristics by incorporating additional metadata signals and full-text cues while maintaining comparable quality controls. Expanding full-text coverage would increase the proportion of reviews for which methodological information can be structured and analyzed. Further work is also needed to approximate or reconstruct included-study sets more directly, enabling evaluation settings that are closer to screening-level gold standards. Finally, improvements in extraction methods that increase recall without compromising evidence grounding would strengthen the completeness of the structured representation while preserving verifiability.




%% file: sigir26-sr4all-sum.tex
\section{Conclusion}

Webis-SR4ALL-26 is a large-scale, cross-disciplinary corpus of systematic reviews comprising 301,871~reviews across 27~scientific disciplines. It integrates resolved OpenAlex reference lists, structured methodological information extracted through a verification-based pipeline, and executable approximations of reported search strategies within a unified open infrastructure. The corpus supports cross-domain benchmarking of retrieval and screening components, training and evaluation of extraction methods for review artifacts, and comparative analyses of systematic review practices across disciplines and time. By enabling reproducible research under consistent cross-disciplinary conditions, Webis-SR4ALL-26 extends existing resources that are typically limited in scale and disciplinary scope, and supports work in information retrieval, evidence synthesis, and meta-research on scientific practice.